\newcommand{\antikt}{anti-$k_t$\xspace}
\newcommand{\mur}{\ensuremath{\mu_{\rm R}}\xspace}
\newcommand{\muf}{\ensuremath{\mu_{\rm F}}\xspace}
\newcommand{\fherwig}{{\sc Herwig~6}\xspace}
\newcommand{\fpythia}{{\sc Pythia~6.4}\xspace}
\newcommand{\powheg}{{\sc Powheg}\xspace}
\newcommand{\mcatnlo}{{\sc MC@NLO}\xspace}
\newcommand{\sherpa}{{\sc Sherpa}\xspace}
\newcommand{\xsec}{{cross section}\xspace}

\newcommand{\xsecs}{{cross sections}\xspace}

\newcommand{\ptjj}{\ensuremath{p_{{\mathrm T},bb}}\xspace}
\newcommand{\mjj}{\ensuremath{m_{bb}}\xspace}
\newcommand{\dr}{\ensuremath{\Delta R}\xspace}
\newcommand{\dphi}{\ensuremath{\Delta \phi}\xspace}

\documentclass[cernpreprint,texlive=2011,txfonts,UKenglish]{latex/atlasdoc}
\pdfoutput=1  
\usepackage{latex/atlaspackage} 
\usepackage{latex/atlasphysics}
\usepackage{latex/atlasbiblatex}
\usepackage{a4wide}
\usepackage{tikz}
\graphicspath{{logos/}{figures/}}

\usepackage{rotating}

\AtlasTitle{Measurement of the $b\overline{b}$ dijet \xsec in $pp$ collisions at $\sqrt{s} = 7$\,TeV with the ATLAS detector.}

\author{The ATLAS Collaboration}

\AtlasRefCode{STDM-2013-03}

\PreprintIdNumber{CERN-EP-2016-091}

\AtlasJournalRef{Eur. Phys. J. C 76 (2016) 670}
\AtlasDOI{10.1140/epjc/s10052-016-4521-y}

\AtlasAbstract{%
The dijet production \xsec for jets containing a $b$-hadron ($b$-jets)
has been measured in proton--proton collisions  
with a centre-of-mass energy of $\sqrt{s} = 7$\,TeV, using the ATLAS
detector at the LHC. The data used correspond to an  
integrated luminosity of $\mathrm 4.2\,\text{fb}^{-1}$. 
The \xsec is measured for events with two identified $b$-jets with a transverse
momentum $\pt > 20$\,GeV and
a minimum separation in the $\eta$--$\phi$ plane of $\Delta R = 0.4$.
At least one of the jets in the event is required to have $\pt > 270$\,GeV.
 The \xsec is measured differentially as a function of dijet invariant mass,  
dijet transverse momentum, boost of the dijet system, and the rapidity
difference, azimuthal angle and angular distance  between the
$b$-jets. 
The results are compared to different predictions of
leading order and next-to-leading order perturbative quantum chromodynamics matrix elements supplemented with models for parton-showers and hadronization.
}

\AtlasCoverSupportingNote{ATL-COM-PHYS-2013-325}{https://cds.cern.ch/record/1528139}
\AtlasCoverCommentsDeadline{11 April 2016}

 \AtlasCoverAnalysisTeam{{Monica Verducci}~~{monica.verducci@cern.ch}\\ {Daniela Salvatore}~~{daniela.salvatore@cern.ch} \\{Fabio Anulli}~~{fabio.anulli@cern.ch}}

\AtlasCoverEdBoardMember{{Ian Brock}~~{brock@physik.uni-bonn.de}}
\AtlasCoverEdBoardMember{{Jean-Francois Arguin}~~{arguin@lps.umontreal.ca}}
\AtlasCoverEdBoardMember{{Caterina Doglioni}~~{Caterina.Doglioni@cern.ch}}
\AtlasCoverEdBoardMember{{Tony Doyle}~~{tony.doyle@glasgow.ac.uk}}

 \AtlasCoverEgroupEditors{atlas-stdm-2013-03-editors@cern.ch}

\AtlasCoverEgroupEdBoard{atlas-stdm-2013-03-editorial-board@cern.ch}

\hypersetup{pdftitle={ATLAS document},pdfauthor={The ATLAS Collaboration}}

\begin{document}

\maketitle

\tableofcontents

\section{Introduction}
\label{sec:intro}

The measurement of jets containing a $b$-hadron ($b$-jets) produced in proton--proton collisions at the Large Hadron Collider (LHC) provides an important test of perturbative quantum chromodynamics (pQCD). Calculations 
of the $b$-quark production \xsec have been performed at next-to-leading order of $\alpha_{\rm s}$ (NLO) in pQCD~\cite{Nason:1987xz,Nason:1989zy,Beenakker:1990maa,Frixione:1997ma}. 
These calculations can be combined with different parton-shower and hadronisation models to generate simulated events which can be compared to data. 

Cross sections for the production of a \bbbar pair have been measured previously at the Tevatron~\cite{Abe:1993sj,Abe:1993hr,Abbott:1999se,Abbott:2000iv}, 
and at the LHC by the ATLAS~\cite{STDM-2010-15,STDM-2011-40} and
CMS~\cite{CMS-BPH-10-010} collaborations.
These measurements agree with NLO predictions for
well-separated $b$-jets, although $b$-jets with large transverse momenta 
in the central regions 
are not well described by simulations~\cite{STDM-2010-15}.
The results in Ref.~\cite{STDM-2011-40} also agree with the NLO predictions;
though small deviations are present at large transverse momenta in events with a $b$-jet and light-flavour
jet (jet generated by a light quark).
The CMS measurement found that in the 
phase-space region of small angular separation between the $b$-jets, 
there are substantial differences between data and NLO
predictions, and among the NLO predictions themselves.

\begin{figure}[htbp]
\begin{center}
  \includegraphics[width=0.5\textwidth]{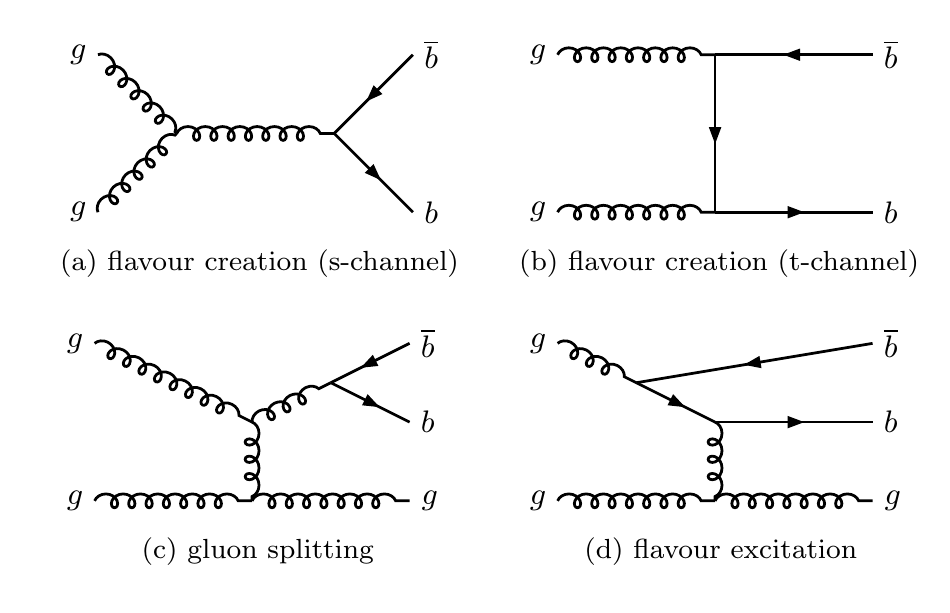}	
    \caption{Lowest-order Feynman diagrams for \bbbar production.}
    \label{fig:bbproduction}
  \end{center}
\end{figure}

The lowest-order Feynman diagrams for \bbbar production are shown in Fig.~\ref{fig:bbproduction}. 
They define different production mechanisms which are useful in understanding the behaviour of the \bbbar system. In 
\emph{flavour creation} (FCR) both $b$-jets originate from the hard scatter: these jets tend to be the hardest in the event and are 
predicted to have an approximate back-to-back configuration in the transverse plane. The \emph{gluon splitting} (GSP) production mechanism creates a pair
of $b$-jets that are  expected to have a small angular separation. The topology of \emph{flavour excitation} (FEX) is less distinctive, 
but it tends to contain an additional parton, which reduces the angular separation between the $b$-jets. 
The requirement of a minimum transverse momentum ($\pt$) of 270\,\GeV\ for the leading jet applied in this
analysis enhances the three-jet production mechanisms relative to the flavour creation mechanism, in comparison to the analyses of
Refs.~\cite{STDM-2010-15,STDM-2011-40,CMS-BPH-10-010}.

Different regions of the \bbbar phase space are probed via the six differential \xsecs presented in this article. 
For large values of the dijet invariant-mass, \mjj, the flavour
creation mechanism is expected to dominate, leading to final states with well-separated hard jets.
Events produced via gluon splitting or flavour excitation are concentrated at small \mjj. The opposite is expected for the \pt\footnote{The ATLAS 
reference system has the origin at the nominal interaction point. The
$x$- and $y$-axes define the transverse plane, the azimuthal angle
$\phi$ is measured around the beam axis, $z$, and the polar angle
$\theta$ with respect to the $z$-axis. The pseudorapidity is defined
as $\eta =  -\ln\left[\tan\left(\theta/2\right)\right]$ and \pt is momentum transverse to $z$.} of the dijet 
system, \ptjj, where the  higher-\ptjj regions are dominated by gluon-splitting production, and only the lower 
values of \ptjj have significant contributions from events produced
via flavour creation. 
The azimuthal angle
between two $b$-jets, $\Delta\phi$, separates the different production
mechanisms more evenly. The angular distance between the two $b$-jets,
$\dr = \sqrt{(\Delta\phi)^2 + (\Delta\eta)^2}$,  is a variable often used
in analyses reconstructing heavy objects decaying into two
$b$-jets. The other two observables are the rapidity difference between the two $b$-jets,  
$y^{*} = \frac{1}{2}|y_1 - y_2|$, where $y_{i}$ is the rapidity of
$b$-jet $i$, and the boost of the dijet system, $y_B =
\frac{1}{2}|y_1+y_2|$. The latter is
related to the momentum of the initial-state partons involved in the
hard scatter and it is therefore sensitive to the parton distribution functions (PDFs). 

The measurement of the \bbbar dijet differential \xsecs is performed
with the ATLAS detector, using proton--proton ($pp$) collisions at a centre-of-mass energy of $\sqrt{s} =
7$\,\TeV. The data were recorded in 2011 and correspond to an
integrated luminosity of  $\mathrm 4.2\, \text{fb}^{-1}$. The differential \xsecs are defined as 
\begin{equation}
\frac{{\mathrm d}\sigma\left(pp \to \bbbar + X\right)}{{\mathrm d}\mathcal{O}} 
  = \frac{N_{\mathrm {\text{tag}}}f_{bb}\ \mathcal{U}}{\varepsilon\mathcal{L}\Delta \mathcal{O} },
\end{equation}
with $\mathcal{O}$ the dijet observable under investigation, $N_{\mathrm {tag}}$ the number of $b$-tagged jet pairs, $f_{bb}$ the purity of the 
selected sample, $\varepsilon$ the selection efficiency, $\mathcal{L}$ the integrated luminosity and $\mathcal{U}$ the correction of the  
measured distribution for detector effects, such as the jet energy resolution. The measurement ranges for the different variables are listed in Table~\ref{tab:fiducial}.

\begin{table}[ht]
\centering
  \vspace{0.5cm}
    \begin{tabular}{c c}
    \hline
    \hline
      Variable & Range \\
      \hline
      \mjj & 50--1000\,\GeV \\
      \ptjj & 0--400\,\GeV \\
      \dphi & 0--$\pi$ \\
      \dr & 0.4--4.0 \\
      $y_B$ & 0--2.5 \\
      $y^*$ & 0--1.7 \\
      \hline
      \hline
    \end{tabular}
    \caption{Ranges of the variables of the measured differential cross sections.} 
    \label{tab:fiducial}
\end{table}

\section{ATLAS detector}
\label{sec:detector}

The ATLAS detector~\cite{PERF-2007-01} 
consists of an inner tracking system, immersed in a 2\,T axial magnetic field, 
surrounded by electromagnetic calorimeters, hadronic calorimeters and a muon spectrometer.

The inner detector has full coverage in $\phi$ and covers the pseudorapidity range $|\eta| < 2.5$. The inner detector consists of 
silicon pixel and microstrip detectors, surrounded by a transition radiation tracker (up to $|\eta| = 2.0$). The electromagnetic 
calorimeter is a lead--liquid argon sampling calorimeter covering $|\eta| < 3.2$. Hadron calorimetry in the central pseudorapidity region 
($|\eta| < 1.7$) is provided by a scintillator-tile calorimeter using steel as the absorber material. The  hadronic end-cap 
calorimeter uses liquid argon with copper absorber plates and extends up to $|\eta| = 3.2$. Additional forward calorimeters 
extend the coverage to $|\eta| < 4.9$. The outer region of the detector is formed by a muon spectrometer that 
uses a toroidal magnetic field with a bending power of 1.5--5.5\,Tm in the barrel and 1.0--7.5\,Tm in the end-caps.
The muon spectrometer provides trigger information for muons up to $|\eta| = 2.4$ and momentum measurements in the bending plane up to $|\eta| = 2.7$. 

The trigger system uses three consecutive  levels to record a selection of interesting events. The level-1 trigger (L1) is 
based on custom-built hardware that processes the data with a fixed latency of 2.5\,${\mathrm\mu}$s. The second level and the 
event filter, collectively referred to as the high-level trigger (HLT), are software-based triggers. 

The jet triggers at L1 use information about the energy deposits in the electromagnetic and hadronic calorimeters using trigger towers with a 
granularity of $\Delta\phi \times \Delta\eta = 0.1 \times 0.1$. Jet identification is based on the transverse energy deposited 
in a sliding window of $4\times 4$ or $8\times 8$ trigger towers. The HLT further refines the selection, making use of finer-granularity
detector information and using reconstruction software close to that used by physics analyses. Due to the high rate of jet production, 
only a predetermined fraction of events that pass the jet triggers are
recorded. The factor by which the number of events that pass a trigger
is reduced is known as the {\textit{prescale}}. 

\section{Simulated dataset}

To investigate efficiencies and model the data, simulated dijet events produced by the \fpythia~\cite{Sjostrand:2006za}  Monte Carlo (MC)
event generator are used. \fpythia implements matrix elements at leading order (LO) in $\alpha_{\rm s}$ for 2$\to$2 processes, 
a \pt-ordered parton shower with 
leading-logarithm accuracy and multi-parton interactions to simulate the underlying event. The hadronisation is described 
using the Lund string model~\cite{Andersson:1983ia}. Events are generated with the 
MRST LO**~\cite{Sherstnev:2008dm} PDFs and a set of parameters tuned to ATLAS data, 
AUET2B-LO**~\cite{ATL-PHYS-PUB-2011-009}. At this stage, all generated particles with a lifetime greater than 30\,ps are 
collectively referred to as the \textit{particle-level}
event. \textit{Detector-level} events are produced by passing the
particle-level events through a full simulation~\cite{SOFT-2010-01}
of the ATLAS detector based on GEANT4~\cite{Agostinelli:2002hh}. 
The effect of multiple $pp$ interactions in the same or nearby bunch crossings (pile-up)
is included in all MC simulations. Events in MC simulation are reweighted, 
in order to match the distribution of the number of multiple $pp$ interaction distributions to that observed in the data. 
During the 2011 data-taking period, the number of $pp$ collisions per bunch crossing varied between
0.5 and 24~\cite{DAPR-2011-01}.
The resulting simulated events are digitised to model the detector responses, and then reconstructed using the same software as for data processing.

\section{Jet selection}

Jets are reconstructed from energy clusters in the calorimeter using the \antikt~\cite{Cacciari:2008gp,Cacciari:2005hq} algorithm 
as implemented in the FastJet package~\cite{Cacciari:2011ma}, with jet radius parameter $R=0.4$. Jet energy is corrected to the 
hadronic energy scale~\cite{PERF-2012-01}, which on average adjusts the reconstructed
jet energy to the true energy.
The reconstructed jets are subjected to calorimeter-based quality
selections~\cite{DAPR-2012-01}.  Jet candidates coming
from background processes, namely: cosmic-ray showers, LHC beam conditions and hardware problems, are rejected as described in
Ref.~\cite{ATLAS-CONF-2010-038}. 
 Central jets with $|\eta|<2.5$
originating from pile-up are rejected by a track-based 
selection~\cite{ATLAS-CONF-2013-083}. 

\subsection{$b$-jet selection}
\label{ref:btagging}

The flavour of a jet at particle level is defined according to the hadrons contained in the jet. If the jet contains at least one $b$-hadron with 
$\pt > 5$\,\GeV\ and \dr with respect to the jet axis of less than 0.3, then it is considered as a $b$-jet. If no $b$-hadron is present, 
but a $c$-hadron that meets the same criteria is found, then the jet is considered as a $c$-jet. All other jets are considered as 
light-flavour jets.

At detector level, the relatively long lifetime of $b$-hadrons is used to select an event sample enriched in $b$-jet pairs. 
To identify $b$-jets, a combination of the JetFitter and IP3D algorithms~\cite{ATLAS-CONF-2011-102} is used. 
The JetFitter algorithm aims at reconstructing the decay vertex of the $b$-hadron 
and the subsequently produced $c$-hadron, assuming that both vertices lie on the same line from the primary vertex,\footnote{The primary
vertex is defined as the vertex with the largest scalar sum of $p_{\mathrm T}^{2}$ for its associated tracks and with at least two associated tracks with $\pt>400$ \,\MeV.} corresponding to the flight direction of the $b$-hadron. The IP3D algorithm is a track-based algorithm using the signed longitudinal 
and transverse impact parameter significances of the tracks matched to the jet (where the impact parameter is defined as the distance of the track from the vertex at the point of closest approach). The variables describing the impact parameters and the 
reconstructed decay chain are combined by a neural network trained
using MC simulation samples. This combination assigns a set of
probabilities $(p_b, p_c, p_l)$ to every jet, corresponding to the
probability of the jet being a $b$-jet, $c$-jet  
or light-flavour jet, respectively. A jet is considered to be $b$-tagged when $\log_{10}\left(p_b/p_l\right) > 0.35$, a 
choice that results in a $b$-tagging efficiency of $\varepsilon_b \sim
70\%$ in simulated \ttbar events (corresponding to a $c$-jet rejection factor of 5 and light-flavour jet rejection factor of 125).

A scale factor is applied to the efficiency obtained from simulation to account for the data-MC difference.
Two methods are employed to select these $b$-jet samples: the first
uses an independent $b$-tagging algorithm that selects jets containing  
a muon from a semileptonic $b$-hadron decay~\cite{ATLAS-CONF-2012-043}; the other selects $b$-jets from \ttbar 
decays~\cite{ATLAS-CONF-2012-097}. The differences between data and simulation observed in these control samples are used to derive a series of 
\pt- and $\eta$-dependent scale factors, which are then applied to each jet in simulation. 

Table~\ref{tab:definition} contains all the fiducial phase-space
definitions for particle-level and detector-level objects used in this
analysis.

\begin{table}[ht]
\centering
  \vspace{0.5cm}
  
    \begin{tabular}{l c c}
    \hline
    \hline
      Definition &Particle-level jets & Detector-level jets \\
      \hline
      Jet identification &  \antikt with $R=0.4$& \antikt with $R=0.4$\\
      & include muons and neutrinos & \\ \hline
      $b$-jets definition & $b$-hadron with $\pt > 5$ \,\GeV &$\log_{10}\left(p_b/p_l\right) > 0.35$ \\
      & $\Delta$R (jet;$b$-hadron)<0.3& \\\hline\hline
      Event selection & &\\ \hline
     Leading jet & \multicolumn{2}{c}{$\pt > 270$\,\GeV\ and $|\eta| < 3.2$} \\ \hline
      2 $b$-jets selection &\multicolumn{2}{c}{$\pt > 20$\,\GeV\ and $|\eta| < 2.5$} \\
      & \multicolumn{2}{c}{2 $b$-jets separated by $\dr~>~0.4$}\\
       \hline
      \hline
    \end{tabular}
  \caption{Fiducial phase space of the measurement. The definition and
    the selection requirements for particle-level  and detector-level
    jets are given. The particle-level jets are constructed using all particles, including muons and neutrinos, as input (see Sect.~\ref{ref:Unfolding}).
    The $\log_{10}\left(p_b/p_l\right)$ criterion corresponds to the
    ratio of the probabilities of the jet being a $b$-jet or
    light-flavour jet. } 
  \label{tab:definition}
\end{table}

\section{Event selection}

Events are selected using two calorimeter-based single-jet triggers with a \pt-thresholds of 180 and 240\,\GeV and $|\eta| < 3.2$.
These thresholds are used to define two ranges for the transverse momentum of
the leading jet where the trigger efficiency is close to 100\%: 
$270 < \pt < 355$\,\GeV\ and $\pt > 355$\,\GeV, respectively.
A prescale factor of 3.5 was applied to the 180 \,\GeV\ threshold
trigger; no prescale factor was applied to the 240 \,\GeV\ threshold trigger.

Quality requirements are applied to ensure that the selected events are well measured. In 
addition to selecting only data from periods in which all sub-detectors
were operating nominally, a veto is applied to reject specific events
in which the calorimeters were suffering from noisy or inactive
regions. 

The leading jet is not required to be identified as a $b$-jet, but the
selected events must have at least two $b$-tagged jets with $\pt > 20$\,\GeV\ and
$|\eta| < 2.5$,  and  the two highest-\pt $b$-tagged jets within the
$|\eta|$ requirement are taken as the dijet pair. To avoid jets with
significant overlap, the two $b$-tagged jets in the pair are also required to be
separated by $\dr~>~0.4$. 

\begin{figure*}[htbp]
  \subfloat[$90\,{\mathrm{\GeV}} < \ptjj < 110\,{\mathrm {\GeV}}$]{
    \includegraphics[width=0.5\textwidth]{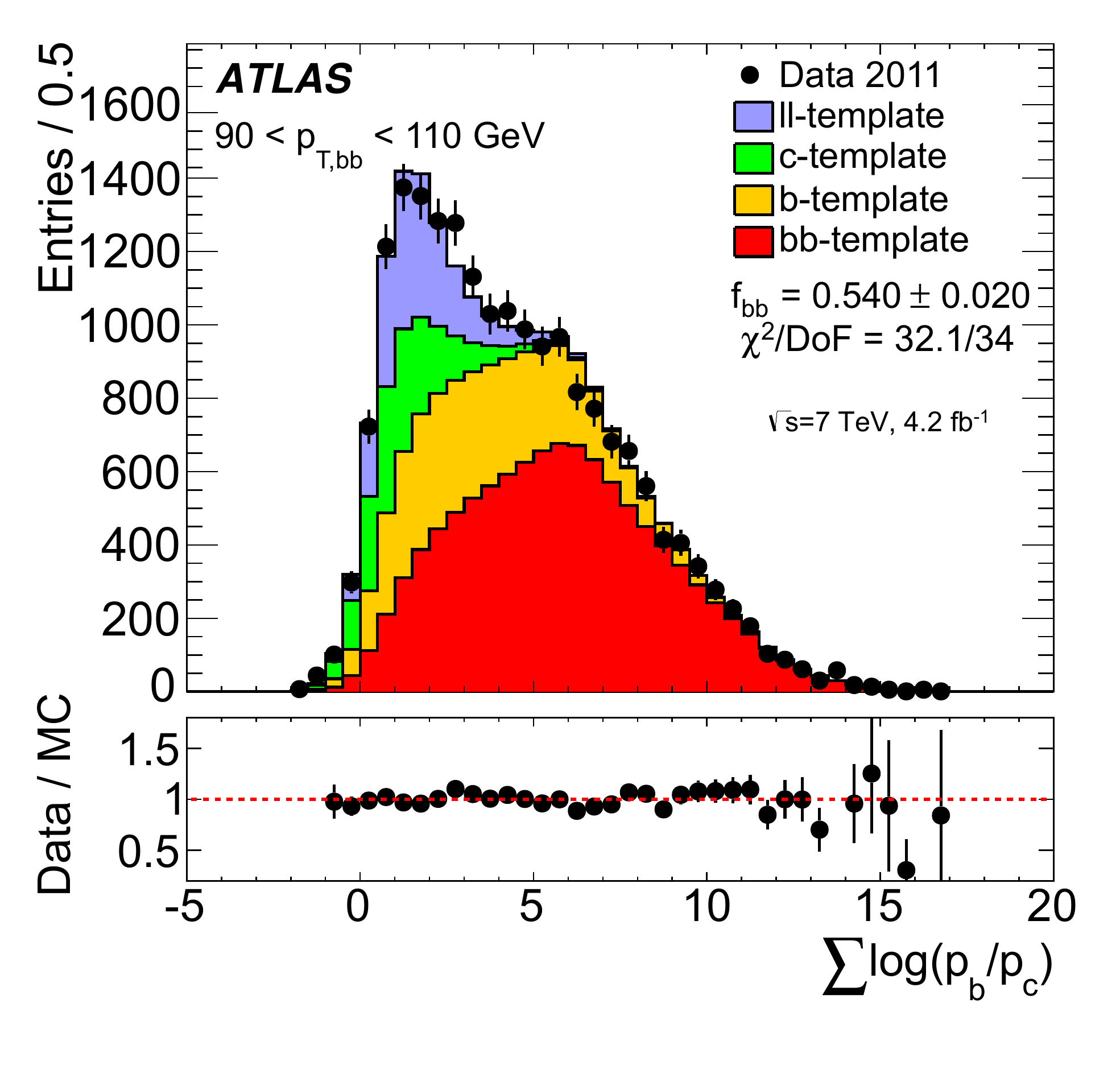}
  }
  \subfloat[$1.4 < \dr < 1.7$]{
    \includegraphics[width=0.5\textwidth]{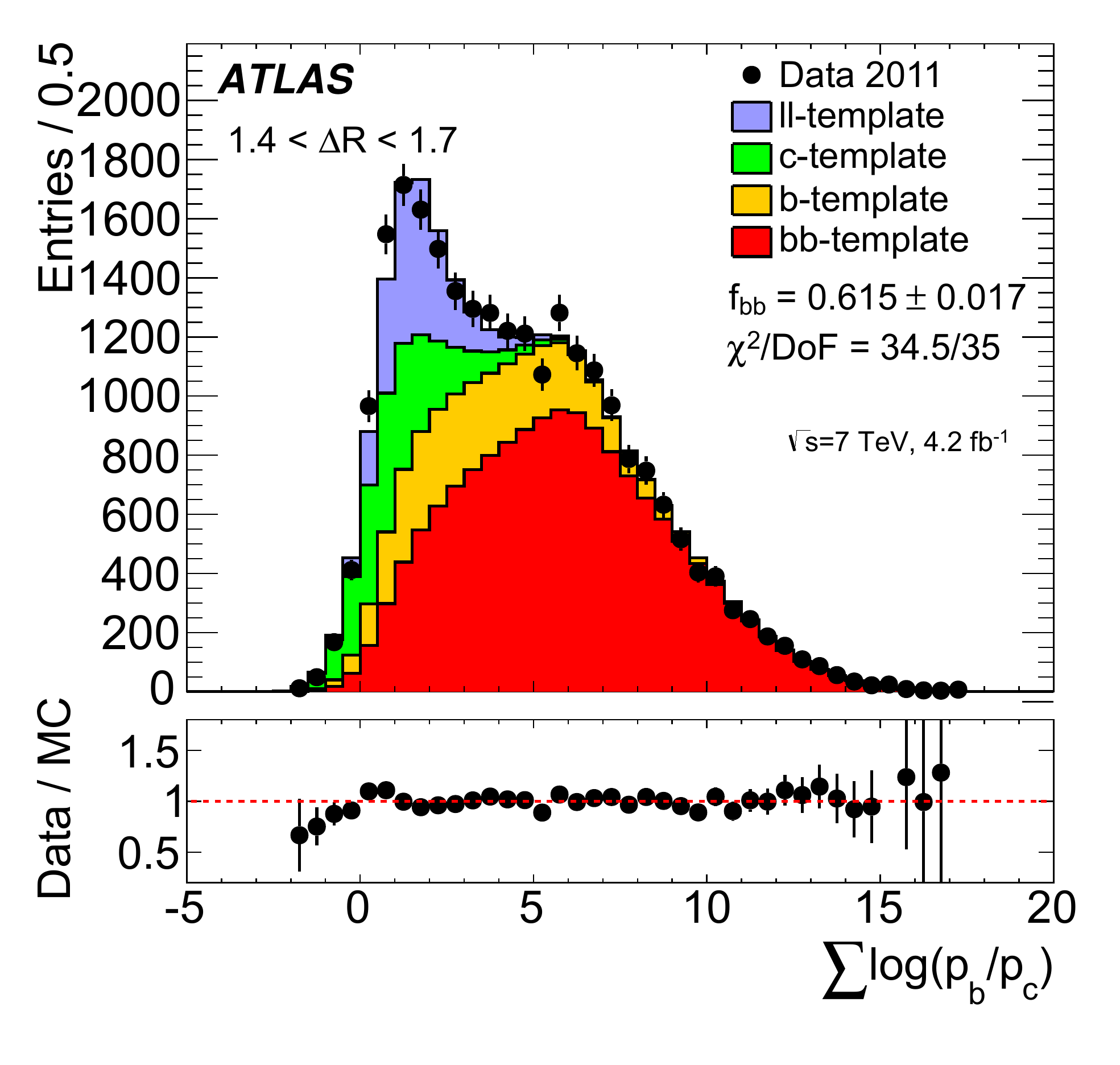}
  }
  \caption{
    Examples of template fits in a \ptjj and \dr bin. The uncertainty
    shown on the \bbbar fraction is the statistical uncertainty 
    of the fit parameter. Systematic uncertainties are not shown.
  }
  \label{fig:fits}
\end{figure*}

\subsection{Purity}
While the requirement of $b$-tagged jets provides an event sample
enriched in \bbbar pairs, there is still a non-negligible contamination from $c$-jets and 
light-flavour jets. The fraction of true $b$-jet pairs in the sample of $b$-tagged jet pairs, referred to as the 
purity of the sample, is determined by performing a template fit to the combined IP3D and JetFitter probability distributions. This fit is performed
independently in each bin of the cross-section measurement. To obtain optimal separation between $b$- and $c$-jets, the fit variable is constructed 
as $\sum \log_{10}\left(p_b/p_c\right)$, where the sum is taken over both of the $b$-tagged jets.

The fit uses a maximum-likelihood method to determine the relative
contributions of four templates that best describe the flavour content of the \bbbar pair in
data. These templates are defined as:  
\begin{itemize}
  \item $bb$-template: $(f_1, f_2)$ = $(b, b)$,
  \item $b$-template: $(f_1, f_2)$ = $(b, c)$, $(c, b)$, $(b, l)$ or $(l, b)$,
  \item $c$-template: $(f_1, f_2)$ = $(c, c)$, $(c, l)$ or $(l, c)$,
  \item $ll$-template: $(f_1, f_2)$ = $(l, l)$,
\end{itemize}
where $f_1$ and $f_2$ indicate the flavour of the leading and sub-leading jet, respectively. The fraction of \bbbar events in the  $b$-tagged sample
is determined by the relative contribution of the $bb$-template. 

The dijet templates are obtained from single-jet templates in MC simulation using a convolution technique in every bin of the investigated variables. This allows 
the creation of smooth, finely binned templates even for bins with a small number of dijet pairs.
As the $b$-hadron decay is a process internal to the jet, the $b$-tagger probabilities for a given jet 
do not depend significantly on the properties of the dijet system. The shape of the fit variable distribution can be parameterised as a function of the \pt and flavour of
the single jets in simulation. The contribution of each $(p_{\mathrm T1}, p_{\mathrm T2}, f_1, f_2)$ combination within a cross-section bin is then determined by 
convolving the $\log_{10}\left(p_b/p_c\right)$ distributions for $(p_{\mathrm
  T1}, f_1)$ and $(p_{\mathrm T2}, f_2)$.
Figure~\ref{fig:fits} shows examples of the fits for two bins of the variables
\ptjj and \dr.

To verify the validity of the procedure, a closure test is performed by comparing the templates obtained via the convolution to those obtained from the bi-dimensional \pt distribution. Good agreement is observed with the generated templates in all kinematic regions.

\section{Unfolding}
\label{ref:Unfolding}

The correction of the measured distribution for detector effects and inefficiencies 
is done via an unfolding procedure that uses the iterative dynamically stabilised (IDS)
method~\cite{Malaescu:2009dm}. 
At particle level, jets are constructed using all particles,
including muons and neutrinos, as input. Particle-level jets are required to  
pass the same kinematic selections as jets reconstructed in the calorimeter. The detector
effects are corrected by using an unfolding matrix, which  
maps the event migrations in a binned distribution from detector level
to particle level. The data are unfolded using the   
unfolding matrix and then compared to the predicted particle-level
distribution. The iterative part of the unfolding allows  
the matrix to be modified to account for mismodelling of the MC
simulation. 
Any statistically significant differences
between  the data and simulation are assumed to originate from 
processes not included in the simulation and are added into the  
unfolding matrix. The data are then unfolded using the modified
unfolding matrix and the process is repeated until no element in  
the unfolding matrix is modified by more than one percent. 
To cross-check the unfolding results, the
unfolding is done with a bin-by-bin method and compared. 
The bin-by-bin method  takes the ratio of the detector-level jet and
particle-level jet distributions, combining all the necessary corrections
into a single factor. It treats each bin as an independent
measurement, behaving as if events appear or disappear within the bin
rather than moving to another. 
The ratio of IDS to bin-by-bin results is about 2\%, except for the mass and the \pt, where differences up to 10\%
were observed. 

The unfolding matrix is derived from simulated \fpythia dijet events,
and is defined for events that have both the particle-level and the   
detector-level jet pairs within the fiducial acceptance of the
analysis. Fiducial and efficiency correction factors are applied to
the  data before and after the unfolding, respectively. 
The fiducial correction, applied before the unfolding, accounts for  
the effects that cause a detector-level jet pair not to be matched
to a particle-level jet pair. The primary reason for the  
mismatch is that one of the particle-level jets has a \pt below the
event selection threshold and so is rejected by the analysis.  
The efficiency corrections, applied after the unfolding, correct
primarily for particle-level jets which meet the event selection criteria
but  are measured outside of the fiducial range at detector level.

\section{Systematic uncertainties }

The systematic uncertainties on the measured cross
sections are evaluated varying the relevant quantities by one standard
deviation, and applying the unfolding procedure; the differences with
respect to the standard  procedure are taken as the uncertainties.
The total systematic uncertainties are obtained by adding the components in quadrature.

The dominant systematic uncertainties in this measurement result from
the $b$-tagging and the jet energy scale
calibrations. Table~\ref{tab:systematics} provides an overview of the
systematic uncertainties. Their magnitude depends on the variable used in the differential cross section measurement, and the minimum and
maximum uncertainties are reported.
Each systematic uncertainty is propagated through the entire analysis,
including the unfolding procedure. 

\begin{table}[ht]
\centering 
  \vspace{0.5cm}
    \begin{tabular}{lr}
    
    \hline
    \hline
      Source & Cross Section Relative Uncertainty \\
      \hline
      $b$-tagging efficiency & 10--30\% \\
      $b$-jet template fit & 3--8\% \\
      Jet energy scale & 10--20\% \\
      Jet energy resolution & 2--8\% \\
      Jet angular resolution & 1--5\% \\
      Unfolding & 5--10\% \\
      Luminosity & 1.8\% \\
      \hline\hline
    \end{tabular}
    \caption{Summary of the dominant sources of systematic uncertainties and their relative effect on the \xsec.}
    \label{tab:systematics}
\end{table}

The $b$-tagging efficiency and light-flavour-jet rejection rates in MC
simulation are calibrated by applying \pt- and $\eta$-dependent scale
factors to the simulated jets~\cite{ATLAS-CONF-2012-097,ATLAS-CONF-2012-043}. These scale 
factors are derived using various data-driven techniques, as  
discussed in Sect.~\ref{ref:btagging}. The uncertainty from this
calibration is evaluated by varying the scale factors for each jet
flavour by one standard deviation. 
The effect of this uncertainty  ranges from 10\% to 20\% in
most bins, and reaches 30\% for low and high values of \mjj.

While the $b$-tagging calibration uncertainty takes into account differences between data and MC simulation in the $b$-tagging efficiency, this does not 
necessarily account for differences in the shape of the $b$-tagger probability distributions. The $b$-tagging algorithm used for this 
analysis makes use of tracks to identify $b$-jets. To account for any effects due to track mismodelling, all the template fits 
are re-evaluated after the simulated events are reweighted to match the small difference in $b$-jet track multiplicity observed in data and the difference is taken
as the uncertainty. The resulting uncertainty in the \xsec amounts to
about 1\%. 
In addition, a 
control sample of $b$-jets is selected using a tag-and-probe method. The data-to-MC ratio of the $\log_{10}\left(p_b/p_c\right)$ for the probe jets 
is fitted with a first-order polynomial. The template fits are then redone after the MC simulation is reweighted 
to match the difference in $\log_{10}\left(p_b/p_c\right)$ seen in
data. The resulting uncertainty is in the range 3$\--8\%$.

The systematic uncertainty resulting from the calibration of the jet
energy scale~\cite{PERF-2012-01} is typically around 10\%, but reaches 20\% for jets with a small angular separation. 
Smaller contributions to the jet uncertainties 
result from mismodelling of the jet energy
resolution~\cite{ATLAS-CONF-2010-054} and  the resolution of the jet direction.  
The uncertainty of the jet energy resolution is estimated by performing an additional Gaussian smearing of the jets by one standard deviation, 
resulting in a 2--8\% uncertainty. 
The uncertainty due to the jet angular resolution is estimated  by
comparing the angular resolution of the nominal sample with
that of samples for which the material description and $b$-jet
fragmentation are varied~\cite{Bowler:1981sb,Buckley:2009bj}; this
uncertainty is in the range 1--5\%.
 
The unfolding uncertainty is evaluated by reweighting the MC simulation
that is used to derive the unfolding matrix to reproduce  
the \xsec measured in data. 
Using the reweighted unfolding matrix results in a 5--10\% change in \xsec, which is assigned as a systematic uncertainty.
Finally, the systematic uncertainty of the luminosity is 1.8\%~\cite{DAPR-2011-01}. 

\section{Theoretical predictions\label{tp}}

The results are compared to the NLO MC generators \powheg, r2299,~\cite{Nason:2004rx,Alioli:2010xd,Frixione:2007vw,Frixione:2007nw} 
and \mcatnlo~4.01~\cite{Frixione:2002ik,Frixione:2003ei}. Both NLO
generators use the CT10~\cite{Lai:2010vv} PDFs 
and a $b$-quark mass of 4.95\,\GeV. Events generated with \powheg are
passed through the \fpythia parton shower and  
\mcatnlo events are showered with
\fherwig.520~\cite{Corcella:2000bw}. 
\fherwig uses an angular ordering parton shower with 
a cluster hadronisation model and employs the MRST LO** PDFs and the set of tuned  parameters AUET2-LO**~\cite{ATL-PHYS-PUB-2011-008}. 
While \powheg and \mcatnlo are both formally accurate to NLO, their treatment of higher-order terms differs. The data are also 
compared to the LO predictions provided by the \sherpa~1.43~\cite{Gleisberg:2008ta} and \fpythia MC generators. \sherpa is capable 
of generating multiple partons in its matrix elements, and was also used to generate \bbbar using a LO 2$\to$3 matrix 
elements for this prediction. As \fpythia is a LO generator, it is not expected to provide an accurate normalisation. 
The \fpythia distributions are normalised to the
integrated \xsec measured in data by applying a factor of 0.61.
\sherpa is found to produce the correct cross-section normalisation.
\powheg{}+\fpythia is chosen as the baseline to examine the
theoretical uncertainties. The largest theoretical uncertainties
derive from the PDF 
uncertainties and uncertainties due to missing higher orders. 
By varying the renormalisation scale, \mur, and the factorisation scale, \muf, which are set to the same value in \powheg,
an estimate of the effects of the missing higher-order terms can be 
made. 
To evaluate the uncertainty, the scales are varied independently 
from one half to twice the central value,
and the cross-section variations are added in quadrature. 
The effect of the scale uncertainties on the NLO prediction ranges from 20\% to  50\%, and dominates the theoretical uncertainty. 
The uncertainties due to the choice of PDFs are estimated from the 52 eigenvectors of the CT10 PDF set evaluated at
68\% confidence level, and are in the range 5--10\% for the variables investigated.
Other cross-checks were performed, such as a study of the
effect due to the $b$-quark mass uncertainty and of the scale matching 
between \powheg and the parton shower. All of these have a negligible effect.
The total theoretical uncertainty is obtained by adding the scale and PDF uncertainties in quadrature.

\section{Results and discussion}
\label{sec:result}

\begin{figure}[htbp]
  \includegraphics[width=\columnwidth]{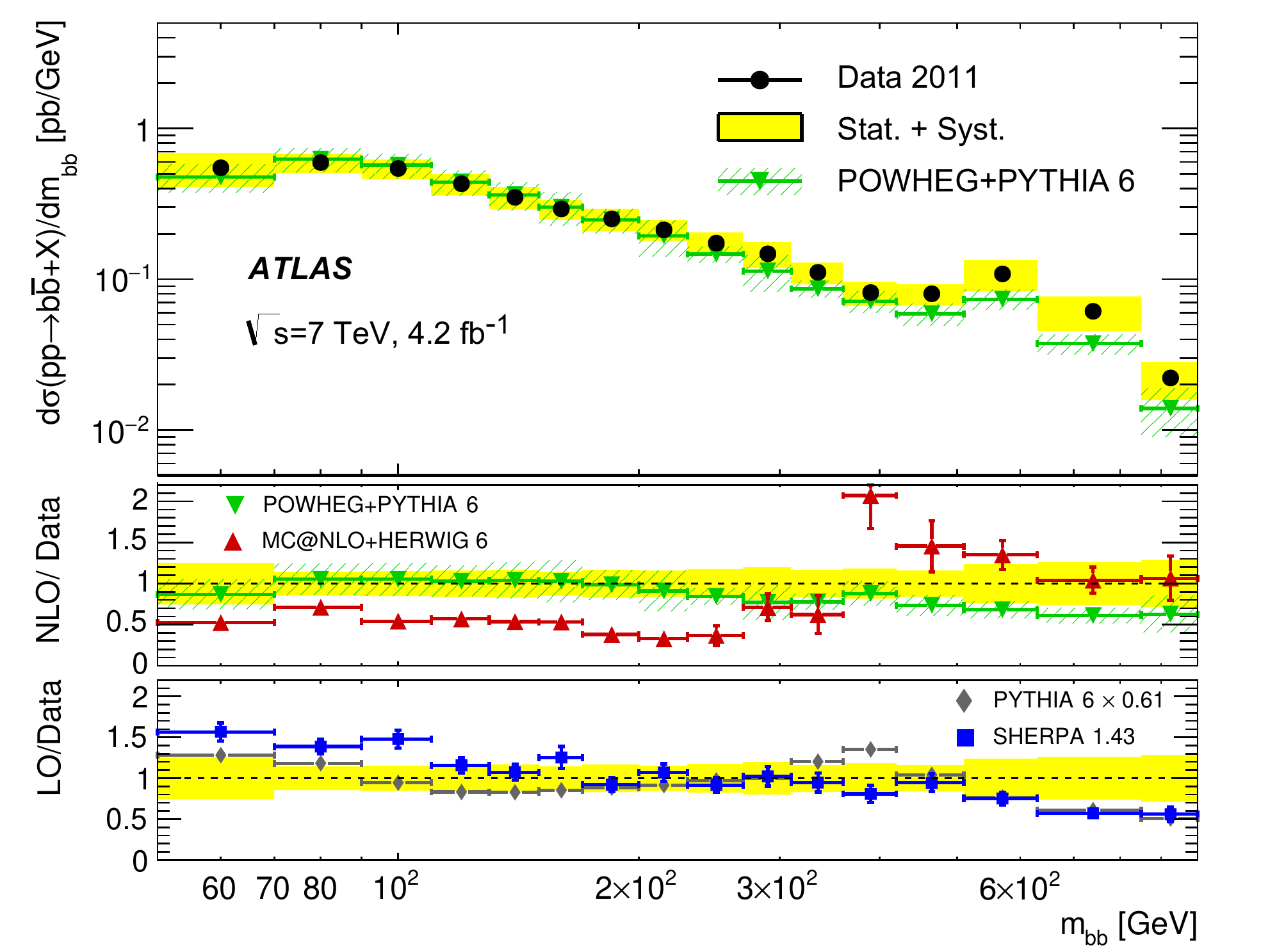}
  \caption{
    Top panel: the differential \xsec for \bbbar production as a function of dijet invariant mass, \mjj, compared to the theoretical 
    predictions obtained using \powheg. Theoretical uncertainties obtained by using
\powheg are also shown. Middle panel:  ratio of the NLO predictions to the 
    measured \xsec. Bottom panel: ratio of the LO predictions to the measured \xsec. For the predictions from \mcatnlo, \sherpa and \fpythia only the statistical uncertainties are shown. For both Middle and Bottom panels: the yellow band represents the combined statistical and systematic experimental errors for the data. Theoretical uncertainties on the POWHEG prediction are also shown.}
  \label{fig:mass}
\end{figure}

\begin{figure}[htbp]
  \includegraphics[width=\columnwidth]{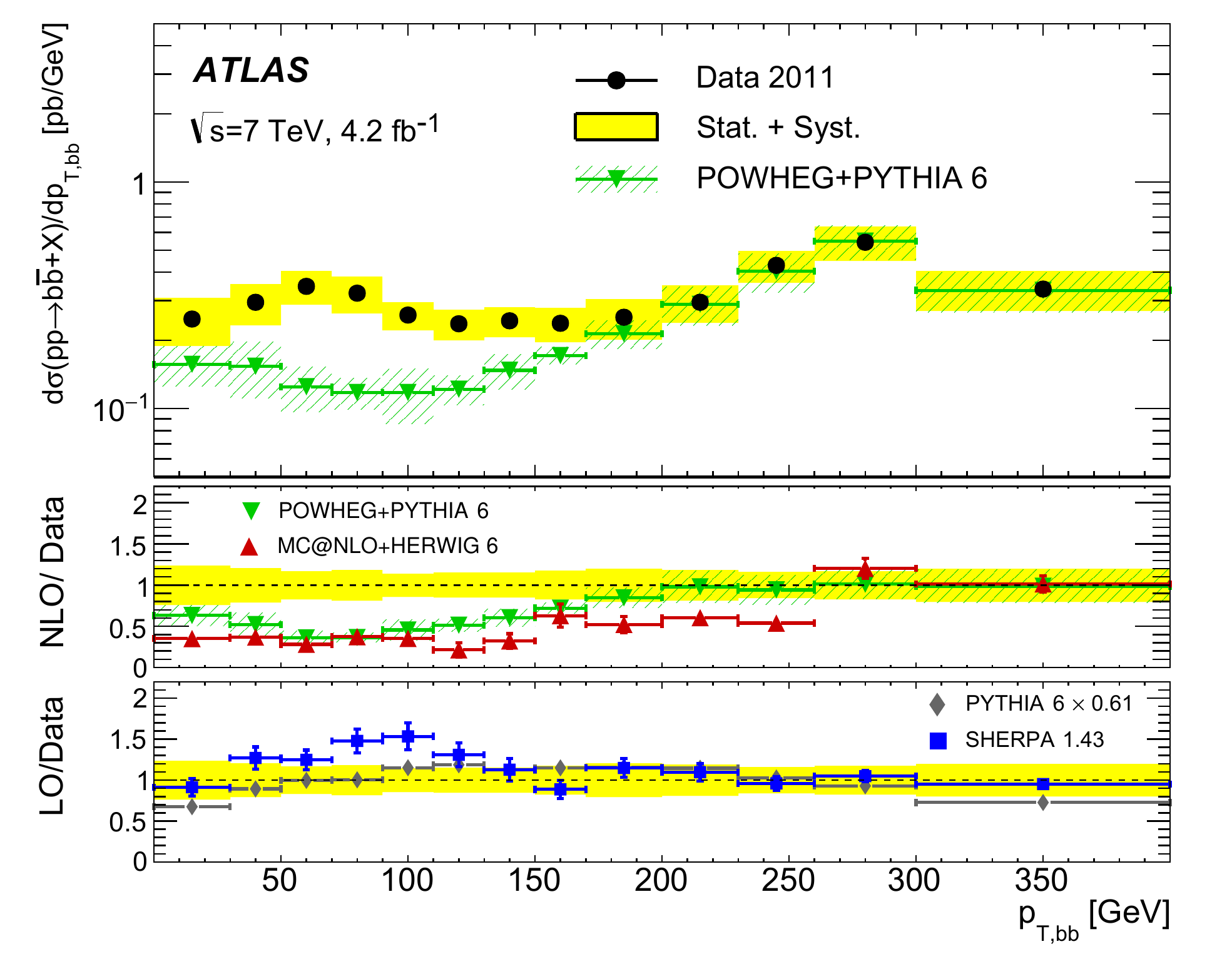}
  \caption{
    Differential \xsec for \bbbar production as a function of the transverse momentum of the dijet system, 
    \ptjj. The figure layout is as in Fig.~\ref{fig:mass}.}
  \label{fig:pt}
\end{figure}

\begin{figure}[htbp]
  \includegraphics[width=\columnwidth]{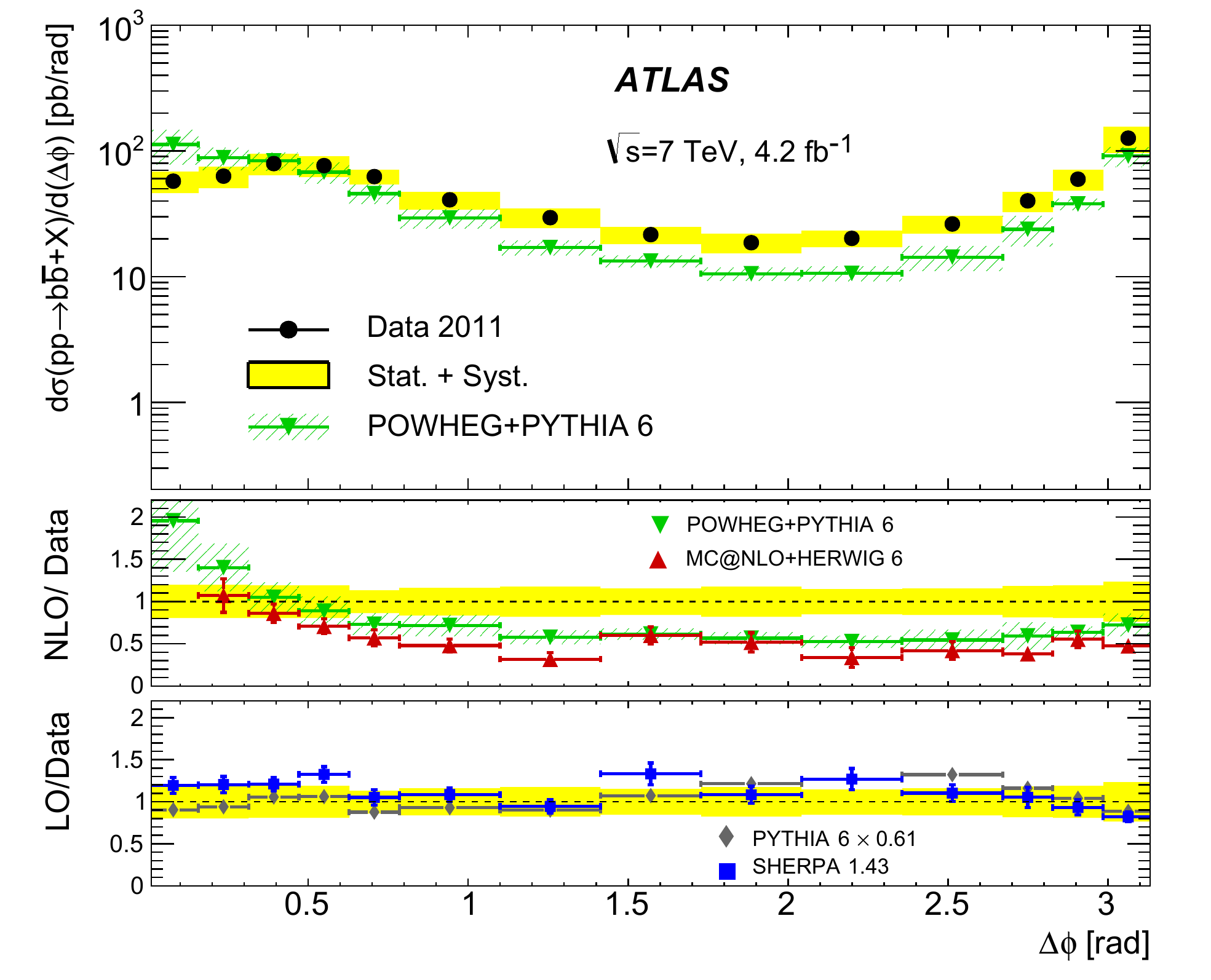}
  \caption{
    Differential \xsec for \bbbar production as a function of the
    azimuthal angle between the two jets, $\Delta\phi$. 
    The point corresponding to the first bin of the \mcatnlo to data
    ratio is about 3.5 and does not appear in the plot. The figure layout is as in
    Fig.~\ref{fig:mass}.}
  \label{fig:dphi}
\end{figure}

\begin{figure}[htbp]
  \includegraphics[width=\columnwidth]{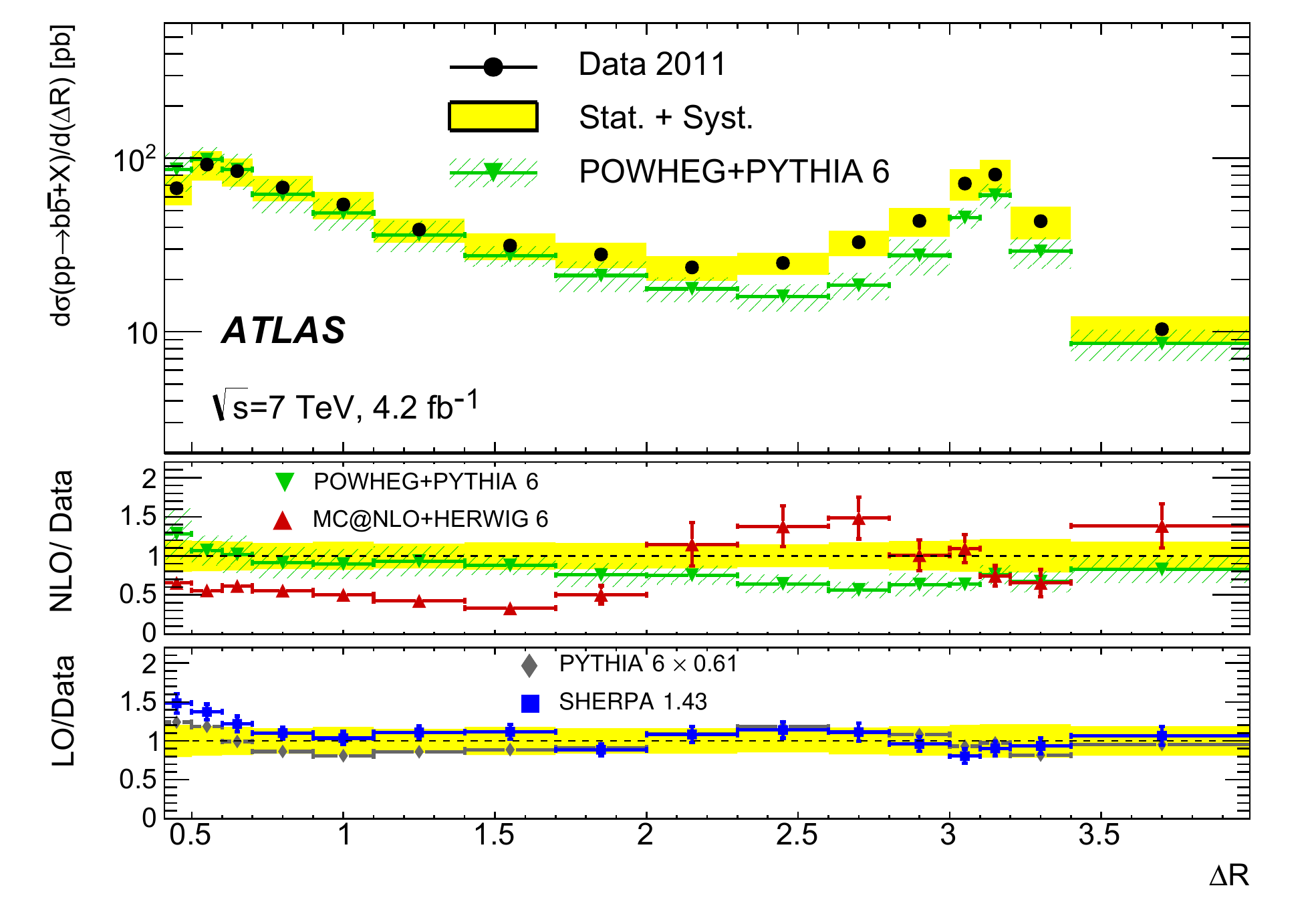}
  \caption{
    Differential \xsec for \bbbar production as a function of the angular distance between the two jets, 
    $\Delta R = \sqrt{(\Delta\phi)^2 + (\Delta\eta)^2}$. The figure layout is as in Fig.~\ref{fig:mass}.}
  \label{fig:dr}
\end{figure}

\begin{figure}[htbp]
  \includegraphics[width=\columnwidth]{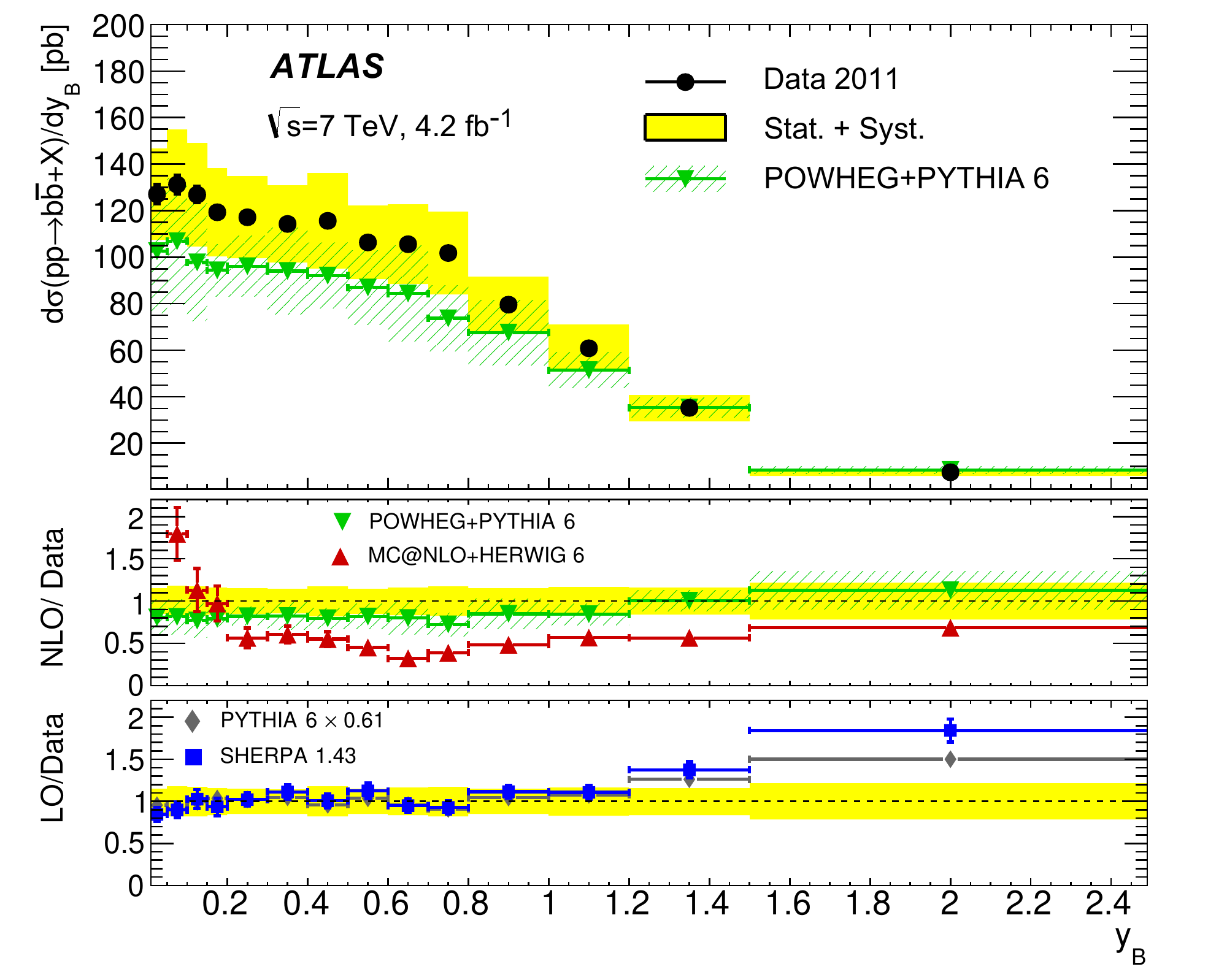}
  \caption{
    Differential \xsec for \bbbar production as a function of the boost of the dijet system, 
    $y_B = \frac{1}{2}\left|y_1+y_2\right|$. The figure layout is as in Fig.~\ref{fig:mass}.}
  \label{fig:yboost}
\end{figure}

\begin{figure}[htbp]
  \includegraphics[width=\columnwidth]{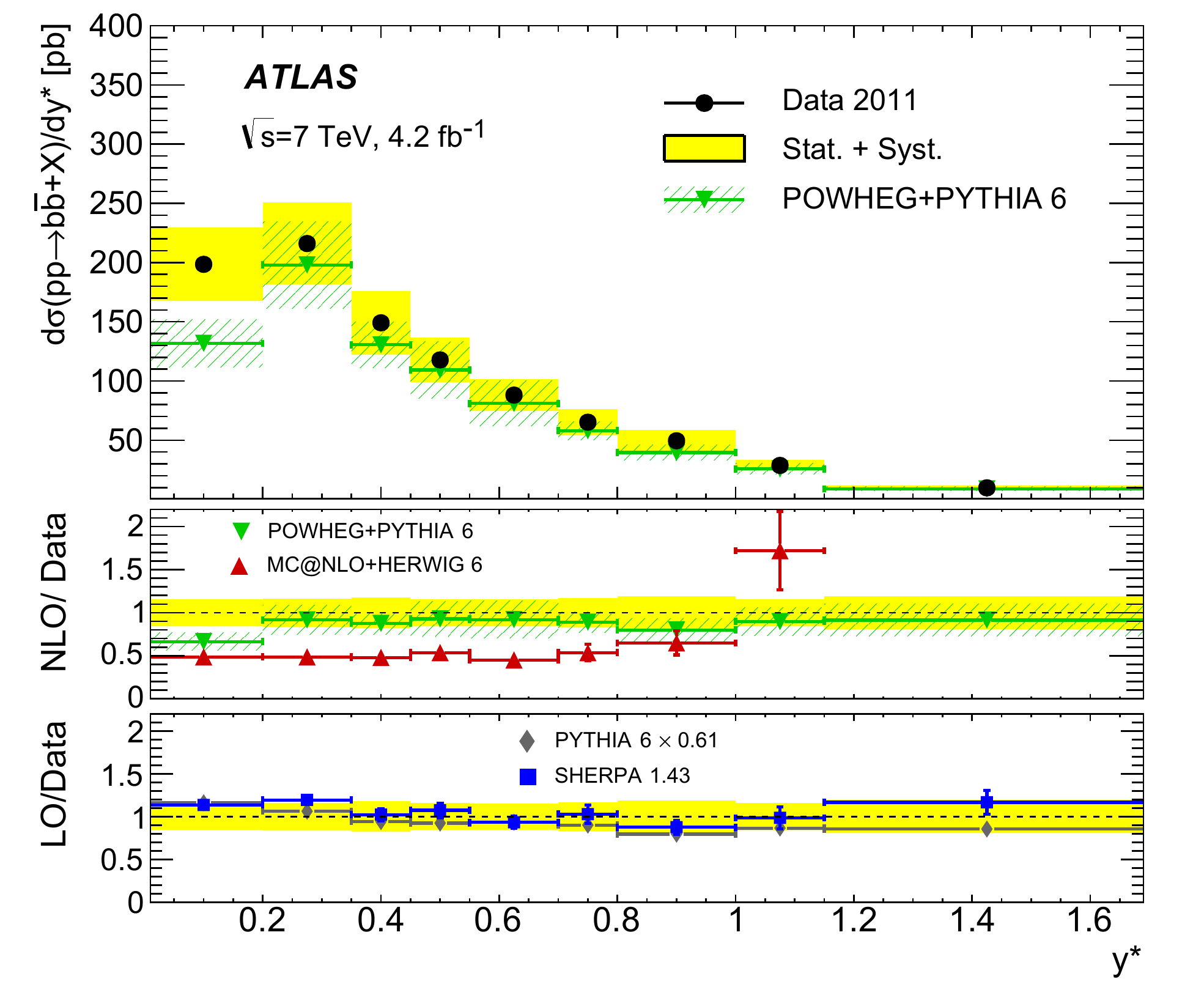}
  \caption{
    Differential \xsec for \bbbar production as a function of $y^{*} = \frac{1}{2}\left|y_1-y_2\right|$. The figure layout is as in 
    Fig.~\ref{fig:mass}.}
  \label{fig:ystar}
\end{figure}

The differential cross section for \bbbar production is shown as a
function of the six observables in Figs.\,\ref{fig:mass}$-$\ref{fig:ystar}. 
The top panel of each figure shows the data points as black dots, with
the total experimental uncertainties as yellow boxes, together with
the prediction and theoretical uncertainties obtained by using
\powheg. 
The middle and lower panels report the ratio of theoretical
predictions to data. 
For the predictions from \mcatnlo, \sherpa and \fpythia, only the statistical uncertainties are shown.
Because of the normalisation factor applied to \fpythia distributions, as explained in Sec.\ref{tp}, the comparison between \fpythia and data is meaningful only at the shape level.
Figure\,\ref{fig:mass} shows the differential cross section for \bbbar
production as a function of the dijet invariant mass.
The cross section decreases with increasing mass except for a step
around 550\,\GeV. This value corresponds approximately to twice the
\pt requirement on the leading jet, i.e. to a mass region where the
flavour-creation process, with two almost back-to-back $b$-jets, becomes the
dominant production mechanism.
\powheg provides a very good description of the data over the whole
mass spectrum, with the exception of the high-mass region where a
small deficit in the prediction is seen.
The \mcatnlo prediction is consistently below the data for 
$\mjj<350$\,\GeV, at which point it becomes higher than data.
This jump corresponds to the region where the flavour-creation process begins
to contribute to \bbbar production. 
The LO predictions (both \sherpa and \fpythia) overestimate the data at low
masses and underestimate them at very high masses. 

The differential cross section as a function of the dijet \pt ranges
between $0.2$ and $0.5 \,\mathrm{pb/\GeV}$, 
as can be seen in Fig.~\ref{fig:pt}. 
Such a relatively constant distribution is a consequence of  requiring a
leading jet with $\pt>270$\,\GeV,  
which suppresses the flavour-creation process, which typically
produces two $b$-jets with low $\ptjj$. 
Without this requirement, flavour creation would overwhelm the
other production mechanisms by several order of magnitudes.
All MC generators provide a good description of the high-$\ptjj$ region.
\powheg and \mcatnlo deviate significantly from data for \ptjj below
about 200\,\GeV, while \sherpa overestimates the data in
the region $50 \lesssim \ptjj \lesssim 130$\,\GeV.
\fpythia reproduces well the shape of the data.

The \xsecs as a function of the azimuthal angle and of the
$\eta$--$\phi$ distance between
the jets are shown in Figs.~\ref{fig:dphi} and \ref{fig:dr},
respectively. 
In these figures, the region at high angular separation is where the
flavour-creation process is expected to dominate. This is visible in
the peaks at $\dphi\sim \pi$ and $\dr\sim 3$.
The NLO predictions are above the data in
Fig.~\ref{fig:dphi} for low \dphi values, where the \bbbar pair is more
likely produced together with at least one other jet. 
They reproduce well the shape of the data distribution for $\Delta\phi \gtrsim 1$,  
but underestimate the cross section by a factor two in the same region.
Good agreement between data and simulation with LO generators
is seen.
The differential cross section as a function of \dr, shown in Fig.~\ref{fig:dr}, is well reproduced
by \powheg.
The ratio of \mcatnlo predictions
to the data is $\sim 0.5$ for $\dr\lesssim 2$, and is above
the data in the intermediate \dr region.
The LO predictions do not show strong
deviations from data apart from an excess for \dr values below $\sim 0.7$.

Figures~\ref{fig:yboost} and \ref{fig:ystar} show the cross sections as a
function of the rapidity variables $y_B$ and $y^{*}$,  
respectively. 
The \powheg predictions reproduce well the shape of the data distribution for
both observables.
The LO predictions deviate from the data for $y_B$ > 1.2.
The \mcatnlo predictions are above the data  for  $y_B < 0.1$,  and are
significantly lower than the data for $ 0.3 \lesssim y_B \lesssim
1.4$, as well as for $y^{*} \lesssim 0.8$.
\fpythia and \sherpa also generally describe the data well, particularly the $y^*$ distribution, 
although their predictions are above the data for $y_B \gtrsim 1.2$.

\section{Conclusion}
\label{sec:conclusion}

Differential \xsecs for pairs of $b$-jets have been measured in $pp$
collisions at $\sqrt{s} = 7$\,\TeV\ using $\mathrm 4.2\,\text{fb}^{-1}$ of data
recorded by the ATLAS detector at the LHC. 
Six dijet variables are investigated to probe the \bbbar phase space:
the invariant mass, 
the transverse momentum, and the boost of the dijet system; 
the azimuthal angle,
the angular separation, 
 and the rapidity difference between the two $b$-jets.
The dijet system is defined as the two highest-\pt $b$-jets
in the event with 
\pt $>$ 20\,\GeV, $|\eta|<2.5$, requiring a minimum \dr of 0.4.
A further requirement of a jet in the event with a minimum transverse
momentum of 270\,\GeV\ is applied.

The results are compared with NLO QCD predictions obtained using
\powheg and \mcatnlo and the LO predictions 
provided by \sherpa and \fpythia. 
The use of single-jet triggers with high $p_{\rm T}$ thresholds significantly changes the relative weight of the different production
processes with respect to an almost unbiased selection~\cite{STDM-2010-15}, with
an enhancement of the gluon-splitting mechanism by strongly suppressing the low-$p_{{\rm T},bb}$ region where the flavour-creation process dominates.
Under these conditions, \mcatnlo shows significant deviations from data for
all variables, both in terms of shape and normalisation. 
\powheg generally reproduces well the  measured differential cross sections, although it 
underestimates the data at low \ptjj.
The LO predictions approximately reproduce all distributions although some bins
show deviations of up to about 50\%.  

In general, this analysis, which is particularly sensitive to the
three-jet topology,  confirms that the current MC  generators have
significant difficulties in describing regions of phase space which are
not dominated by two hard  $b$-jets. 

\bigskip

\section*{Acknowledgements}

We thank CERN for the very successful operation of the LHC, as well as the
support staff from our institutions without whom ATLAS could not be
operated efficiently.

We acknowledge the support of ANPCyT, Argentina; YerPhI, Armenia; ARC, Australia; BMWFW and FWF, Austria; ANAS, Azerbaijan; SSTC, Belarus; CNPq and FAPESP, Brazil; NSERC, NRC and CFI, Canada; CERN; CONICYT, Chile; CAS, MOST and NSFC, China; COLCIENCIAS, Colombia; MSMT CR, MPO CR and VSC CR, Czech Republic; DNRF and DNSRC, Denmark; IN2P3-CNRS, CEA-DSM/IRFU, France; GNSF, Georgia; BMBF, HGF, and MPG, Germany; GSRT, Greece; RGC, Hong Kong SAR, China; ISF, I-CORE and Benoziyo Center, Israel; INFN, Italy; MEXT and JSPS, Japan; CNRST, Morocco; FOM and NWO, Netherlands; RCN, Norway; MNiSW and NCN, Poland; FCT, Portugal; MNE/IFA, Romania; MES of Russia and NRC KI, Russian Federation; JINR; MESTD, Serbia; MSSR, Slovakia; ARRS and MIZ\v{S}, Slovenia; DST/NRF, South Africa; MINECO, Spain; SRC and Wallenberg Foundation, Sweden; SERI, SNSF and Cantons of Bern and Geneva, Switzerland; MOST, Taiwan; TAEK, Turkey; STFC, United Kingdom; DOE and NSF, United States of America. In addition, individual groups and members have received support from BCKDF, the Canada Council, CANARIE, CRC, Compute Canada, FQRNT, and the Ontario Innovation Trust, Canada; EPLANET, ERC, FP7, Horizon 2020 and Marie Sk{\l}odowska-Curie Actions, European Union; Investissements d'Avenir Labex and Idex, ANR, R{\'e}gion Auvergne and Fondation Partager le Savoir, France; DFG and AvH Foundation, Germany; Herakleitos, Thales and Aristeia programmes co-financed by EU-ESF and the Greek NSRF; BSF, GIF and Minerva, Israel; BRF, Norway; Generalitat de Catalunya, Generalitat Valenciana, Spain; the Royal Society and Leverhulme Trust, United Kingdom.

The crucial computing support from all WLCG partners is acknowledged
gratefully, in particular from CERN and the ATLAS Tier-1 facilities at
TRIUMF (Canada), NDGF (Denmark, Norway, Sweden), CC-IN2P3 (France),
KIT/GridKA (Germany), INFN-CNAF (Italy), NL-T1 (Netherlands), PIC (Spain),
ASGC (Taiwan), RAL (UK) and BNL (USA) and in the Tier-2 facilities
worldwide.

\clearpage

\printbibliography

\newpage 
\begin{flushleft}
{\Large The ATLAS Collaboration}

\bigskip

M.~Aaboud$^\textrm{\scriptsize 136d}$,
G.~Aad$^\textrm{\scriptsize 87}$,
B.~Abbott$^\textrm{\scriptsize 114}$,
J.~Abdallah$^\textrm{\scriptsize 65}$,
O.~Abdinov$^\textrm{\scriptsize 12}$,
B.~Abeloos$^\textrm{\scriptsize 118}$,
R.~Aben$^\textrm{\scriptsize 108}$,
O.S.~AbouZeid$^\textrm{\scriptsize 138}$,
N.L.~Abraham$^\textrm{\scriptsize 150}$,
H.~Abramowicz$^\textrm{\scriptsize 154}$
% ...
\bigskip
\\
% ...
$^{12}$ Institute of Physics, Azerbaijan Academy of Sciences, Baku, Azerbaijan\\
% ...
$^{65}$ University of Iowa, Iowa City IA, United States of America\\
% ...
$^{87}$ CPPM, Aix-Marseille Universit{\'e} and CNRS/IN2P3, Marseille, France\\
% ...
$^{108}$ Nikhef National Institute for Subatomic Physics and University of Amsterdam, Amsterdam, Netherlands\\
% ...
$^{114}$ Homer L. Dodge Department of Physics and Astronomy, University of Oklahoma, Norman OK, United States of America\\
% ...
$^{118}$ LAL, Univ. Paris-Sud, CNRS/IN2P3, Universit{\'e} Paris-Saclay, Orsay, France\\
% ...
$^{136}$ $^{(a)}$ Facult{\'e} des Sciences Ain Chock, R{\'e}seau Universitaire de Physique des Hautes Energies - Universit{\'e} Hassan II, Casablanca; $^{(b)}$ Centre National de l'Energie des Sciences Techniques Nucleaires, Rabat; $^{(c)}$ Facult{\'e} des Sciences Semlalia, Universit{\'e} Cadi Ayyad, LPHEA-Marrakech; $^{(d)}$ Facult{\'e} des Sciences, Universit{\'e} Mohamed Premier and LPTPM, Oujda; $^{(e)}$ Facult{\'e} des sciences, Universit{\'e} Mohammed V, Rabat, Morocco\\
% ...
$^{138}$ Santa Cruz Institute for Particle Physics, University of California Santa Cruz, Santa Cruz CA, United States of America\\
% ...
$^{150}$ Department of Physics and Astronomy, University of Sussex, Brighton, United Kingdom\\
% ...
$^{154}$ Raymond and Beverly Sackler School of Physics and Astronomy, Tel Aviv University, Tel Aviv, Israel\\
% ...
$^{a}$ Also at Department of Physics, King's College London, London, United Kingdom\\
$^{b}$ Also at Institute of Physics, Azerbaijan Academy of Sciences, Baku, Azerbaijan\\
$^{c}$ Also at Novosibirsk State University, Novosibirsk, Russia\\
$^{d}$ Also at TRIUMF, Vancouver BC, Canada\\
$^{e}$ Also at Department of Physics {\&} Astronomy, University of Louisville, Louisville, KY, United States of America\\
% ...
\end{flushleft}


\end{document}